# Anisotropy of zero-resistance states in InN films under an in-plane magnetic filed

Xiaowei He[1], Yanhua Dai[1], Ivan Knez[1], Rui-Rui Du[1,3*] Xingqiang Wang[2], Bo Shen[2]

[1]Department of Physics and Astronomy, Rice University, Houston, Texas 77005, USA

[2]State Key Laboratory of Artificial Microstructure and Mesoscopic Physics, School of Physics, Peking University, Beijing 100871, China

[3]International Center for Quantum Materials, Peking University, Beijing 100871, China

## Abstract

We report low temperature current-voltage measurements on n-type InN films grown by molecular beam epitaxy. The zero-resistance state with a large critical current around 1 mA has been observed at 0.3 K. Under in-plane field configuration, the zero-resistance state shows a large anisotropy in critical current for B parallel and perpendicular to applied current. The anisotropic parameter $\gamma = I_{\parallel}/I_{\perp}$ can be up to 2.5 when B = 0.15T. The anisotropy is explained by the vortex flow in the context of type II superconductivity. We have thus established an important aspect of the phenomenology of superconductivity in an otherwise typical narrow gap semiconductor.

Indium nitride, a member of the group III-nitride semiconductors, has recently attracted much attention due to its updated narrow band gap value (0.7 ~ 0.8 eV), revised from previous values of 1.8 ~ 2.1 eV [1]. This revision extends the band gap of the ternary alloy InGaN from the near infrared of InN to the ultraviolet of GaN, therefore allowing the entire optical window to be covered by a single material system [2]. One important property for InN is the existence of an intrinsic electron accumulation layer near the surface of thin films [3, 4]. Due to the electron accumulation layer, metal-InN contacts can always form Ohmic contacts without any annealing [5]. Another interesting property for InN film is the superconductivity at very low temperature which was first reported by Inushima *et al* [6]. The combination of a two-dimensional (2D) electron layer and superconductivity makes InN a promising candidate for novel device applications such as electron-spin injection, and superconductor-semiconductor hybrid structure. Recently, superconducting Nb/InN-nanowire/Nb junctions with large critical currents up to 5.7A have been realized experimentally, demonstrating a good coupling of the semiconductor to the superconductor [7].

Concerning the superconductivity of InN, several groups have reported its existence in n-type InN films [8-11]. Recently, the superconductivity of p-type InN was reported with a transition onset temperature $T_{c,onset}$ around 3.9 K [12]. Moreover, the Meissner effect for InN films has been observed in low temperature ac-susceptibility measurement [13]. From previous reports, the superconductivity of InN has following properties. Firstly, there is an optimum carrier density for the occurrence of superconductivity in InN [8]. The lowest $n_e$ is limited by the Mott transition at ~$10^{17}$ cm$^{-3}$, and the highest $n_e$ is limited by the disorder-induced superconductor to metal or insulator transition at ~$5 \times 10^{20}$ cm$^{-3}$ [14]. Secondly, superconductivity in InN is anisotropic and type II [8, 10, 12]. The critical magnetic field parallel to the a-b plane is much larger than that perpendicular to the a-b plane. In addition, the resistive transition shows a large broadening where the interval between the onset temperature $T_{c,onset}$ and the zero resistance temperature $T_{c,0}$ can be up to 1 K.

To date, no consensus on the superconducting mechanism of InN has been emerged. Some reports consider it as an extrinsic property originated from metallic Indium impurities introduced by the growth process [11], while others believe it is an intrinsic property of InN itself [8, 10, 13]. From the intrinsic mechanism point of view, it is proposed that the crystal lattice of InN contains effective "indium thin films" which are formed by the interaction among the second nearest neighbor In-In atoms. These effective "indium thin films" makes Indium a quasi two-dimensional superconductor, accounting for the observed anisotropy of the critical field [13]. Moreover, the observed broadening of resistive transition is attributed to thermally activated flow of vortices in the context of type II superconductivity [8, 10]. Here, we report low temperature current-voltage measurements on n-type InN film for B parallel to the a-b plane. The experimental data demonstrate a new type of anisotropy which is associated with the zero-resistance states. This kind of anisotropy could provide experimental evidence for the vortices motion in the *superconducting transition* of InN.

Un-doped InN films were grown by MBE on c-sapphire substrates with an approximately 500 nm thick GaN buffer layer; detailed growth information can be found elsewhere [15]. For the sample studied here, thickness of the InN film was 2.4 μm, electron density and mobility was $5.6 \times 10^{17}$ cm$^{-3}$ and 2400 cm$^2$/V.s, respectively. The measurements were performed in a $^3$He refrigerator equipped with a two-axis Nb superconducting magnet providing a perpendicular field

$B_z$ (up to 3T) and a parallel field $B_{//}$ (up to 2T). A square (3 mm × 3 mm) Van de Pauw sample was used where eight Ohmic contacts (pure Indium) were placed around the perimeter.

The longtitudinal resistance along applied transport current was measured as a function of temperature (T) at zero magnetic field, as shown in Fig. 1(a). A dramatic reduction of resistance can be observed, indicating a phase transition from the normal state to the superconducting state. The phase transition occurs at $T_{c,onset}$ ~2.3 K, and reaches zero resistance at $T_{c,zero}$ ~1.4 K. It's worth to mention that the transition temperature of In metal is $T_c$ ~3.4 K. The obvious difference between $T_{c,onset}$ in the InN film and the $T_c$ of In metal suggests that the observed superconductivity cannot be obviously attributed to the Indium impurities introduced by growth process, at least for the case of Indium clusters.

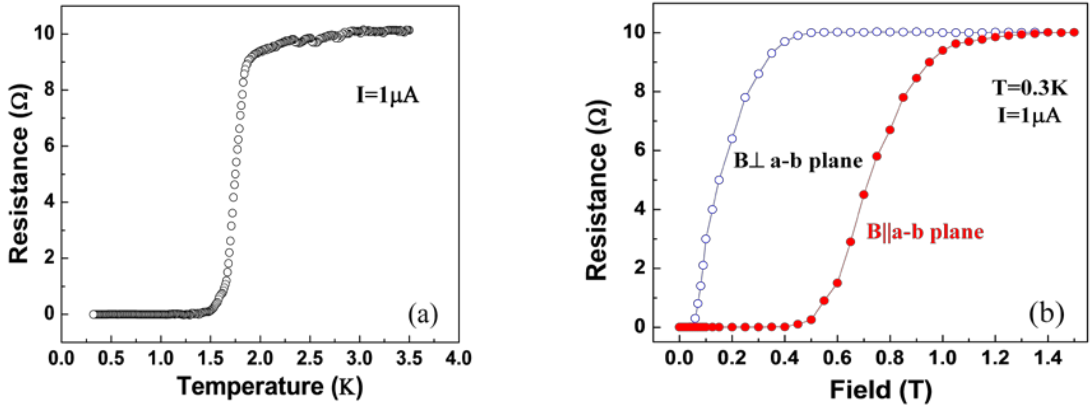

FIG. 1. (Color online) (a) The temperature dependence of longitudinal resistance. (b) magnetoresistance for B perpendicular (blue open circle) and parallel (red solid circle) to a-b plane measured at applied current I = 1µA.

Four terminal magnetoresistance measurements were performed at 0.3 K for both B parallel and perpendicular to the a-b plane. Experimental data showed that the superconductivity in our sample was anisotropic and type II which confirmed previous reports. Generally, for a type

II superconductor there are two critical values of the applied magnetic field for the destruction of superconductivity, which are denoted by $H_{c1}$ and $H_{c2}$. From Fig. 1(b), $H_{c1}$ and $H_{c2}$ can be clearly identified for both B parallel and perpendicular to the a-b plane. They are 0.45 T and 1.35 T for B ∥ a-b plane, and 0.05 T and 0.38 T for B ⊥ a-b plane. The upper critical field $H_{c2}$ for parallel configuration is much larger than that for perpendicular configuration, demonstrating that the superconductivity in our sample is anisotropic (with respect to in- and out of- plane). Moreover, they are much bigger than the critical field of In metal which is about 0.023T, suggesting again that the origin of superconductivity probably cannot be from the presumed In metal, but InN itself.

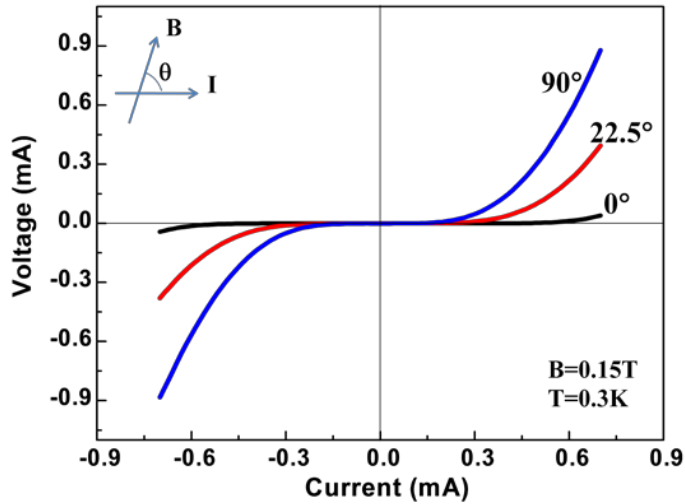

FIG. 2. (Color online) Angular dependence of I-V curve with in-plane magnetic field of 0.15T measured at 0.3 K .

Now we turn to our main experimental findings. The current-voltage (I-V) measurements were performed under an in-plane magnetic field at 0.3 K. By changing the field direction with respect to the applied current, the anisotropy of the zero resistance state was observed at B = 0.1T - 0.5 T. A typical I-V curve with in-plane field B = 0.15 T is shown in Fig. 2. The angle between the direction of in-plane filed and applied current is labeled by θ. It can be found that all I-V curves are strongly nonlinear, indicating the existence of a zero-resistance state with a large critical current. The value of critical current is up to 0.5 mA when in-plane field parallel to applied current (0 degree). But it is dramatically reduced to 0.35 mA for θ = 22.5°, and only around 0.2 mA for θ = 90°, demonstrating a large anisotropy for parallel and perpendicular situations. Defining the anisotropic parameter $\gamma = I_\parallel / I_\perp$, we can find that γ is about 2.5 for B = 0.15 T. It is necessary to note that because the in-plane field is mechanically fixed in the $^3$He refrigerator, experimentally we changed filed direction with respect to applied current by rotating the sample (in a top-loading probe) as a whole. The negligible temperature fluctuation (< 30 mK) induced by rotation was monitored through temperature sensors. Each measurement was performed after the temperature recovered to the same steady value ~ 0.3 K. In addition, the advantage of rotating the whole sample is that we can keep the current direction with respect to the sample surface unchanged, avoiding the possible influences from the anisotropy of lattice structure itself.

We now analyze the mechanism of observed anisotropy in the context of type II superconductivity. Between the Meissner and the normal phase, there should exist a phase called Shubnikov phase or mixed states for type II superconductor [16-18]. In the mixed state region, magnetic field penetrates the superconductor in the form of quantized flux lines, in other words, vortices. Each vortex has a quantized magnetic flux $\Phi_0 = hc/2e$. Applying an external current density $j$ to the vortex system, the flux lines start to move with action of the Lorentz force $F_L = j \times B / c$ which is balanced by the friction force $F_\eta = -\eta v$, where $v$ is the steady state velocity of the vortex, $v = j \times B / c\eta$, and $\eta$ is viscous drag coefficient. The vortex motion produces an electric field $E = B \times v / c$ along the applied current density $j$. This electric field will destroy zero-resistance states and induce power dissipation [17, 18]. Usually, flux lines are pinned by various mechanisms such as lattice defects, yielding a pinning force $F_{Pin}$ which means only for $F_L > F_{Pin}$, vortices start moving [19]. So even in the mixed states region, there exists a critical current density $j_c$.

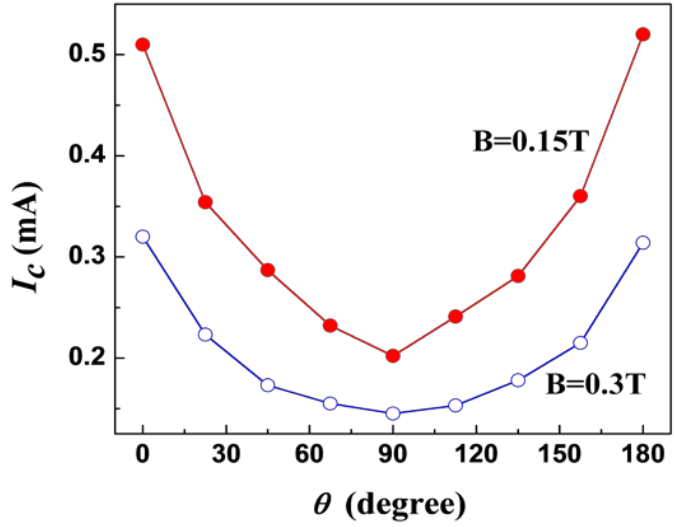

FIG. 3. (Color online) Angular dependence of critical current $I_c$ for in-plane magnetic field B = 0.15T (red solid circle) and B = 0.3T (blue open circle) measured at 0.3K.

From above analysis, it can be found that the flow of vortices is $\theta$-dependent due to Lorentz force $F_L$, and therefore the critical current density $j_c$ is $\theta$-dependent. Coming back to our experiment, for the in-plane configuration the induced electrical field along current direction is obtained as $E \sim jB^2 \sin^2\theta/\eta c$. A quantitative inference about the relationship between critical current and $\theta$ can be drawn from the flow of vortices. Theoretically, when $\theta = 0°$, the Lorentz force $F_L$ acting on flux lines is zero, and the movement of vortices is forbidden. Consequently, a larger critical current is needed to destruct the zero-resistance state. When $\theta = 90°$, the Lorentz force $F_L$ is maximal, the movement of vortices will build a large longitudinal electric field which can easily destroy the zero-resistance state, yielding a reduced critical current. In Fig. 3, we systematically demonstrates the change of critical current as a function of angle θ for B = 0.15T and B = 0.3T. It can be found that critical current $I_c$ changes *periodically* with the change of θ, reaching maxima and then minima from 0° to 90°. Due to that electrical field $E \sim jB^2 \sin^2 \theta / \eta c$, two reversed magnetic fields are equivalent with each other, so when $\theta = 180°$, $I_c$ recovers its initial value.

The above analysis is built on the assumption that the measured sample is in the mixed state. In other words, the applied field should be larger than $H_{c1}$ and smaller than $H_{c2}$. In our experiment, the anisotropy of critical current was observed under a field range from 0.1 to 0.5 T, but from Fig. 1(b) the measured $H_{c1}$ and $H_{c2}$ were 0.45 T and 1.35 T under in-plane configuration. It yields an apparent contradiction since vortex flow cannot occur below $H_{c1}$. This contradiction actually arises from the difference of applied transport current between two kinds of measurement. The applied current in the I-V measurements was up to 0.7 mA, greatly larger than that used in magnetoresistance measurement which was only 0.001 mA. The large transport current could lead to electron heating, and thus the reduction of critical field. To clarify this, we have re-measured the magnetoresistance for in-plane configuration with a high current of 0. 5 mA, as shown Fig. 4(a). Here, indeed, we observed a large reduction of critical fields due to electron heating, in which $H_{c1}$ shrinks remarkably to less than 0.1 T. This result confirms that the sample does enter into the mixed state in the I-V measurements.

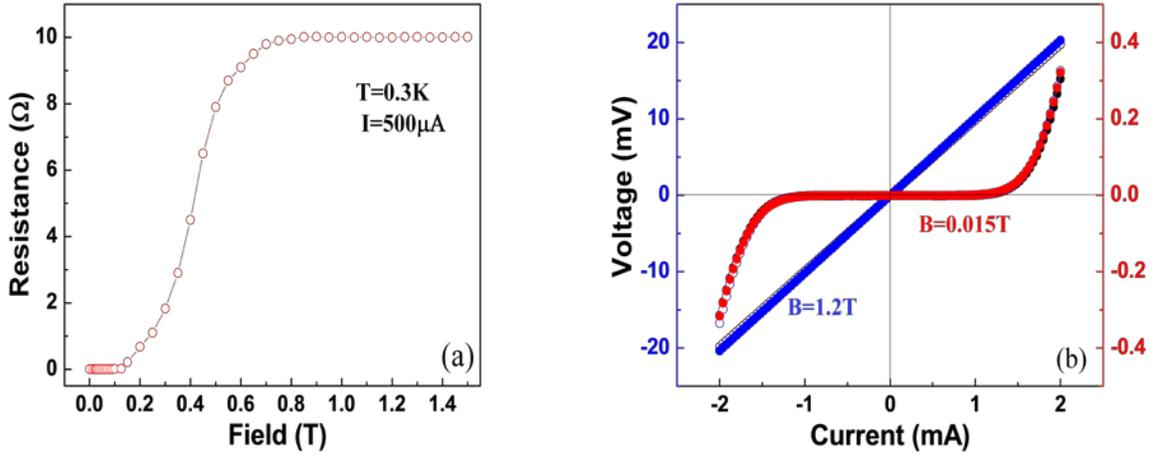

FIG. 4. (Color online) (a) Magnetoresistance for magnetic field B parallel to a-b plane with applied current I = 0.5 mA. (b) Angular dependence of I-V curves for in-plane magnetic field B = 0.015 T and B = 1.2 T, angle θ = 0°, 45°, 90° respectively.

Generally, for the field below $H_{c1}$ or beyond $H_{c2}$ the system enters the Meissner phase or the normal phase, respectively. In either phase, vortices do not exist and thus no anisotropy can be observed in the I-V curve. In order to test this inference, I-V measurements below $H_{c1}$ or beyond $H_{c2}$ were performed. Fig. 4(b) shows the results for B = 0.015 T and 1.2 T. We can see that for B = 0.015 T the critical current is up to 1 mA, but does not show any anisotropy from 0° to 90°. On the other hand, when B = 1.2 T all I-V curves become linear and coincide with each other for different angles. These results are consistent with the vortex flow mode. We thus interpret the observed anisotropy of the zero-resistance states as a strong evidence for vortex motion in the *superconducting transition* of InN.

In conclusion, we have presented current-voltage measurements on a n-type InN film at temperature ∼ 0.3 K. The zero-resistance state with large critical currents around 1mA was observed. Under a small in-plane magnetic field, the zero-resistance state was responsive to the field direction with respect to the applied transport current. The ratio of critical current between B parallel and perpendicular to the applied current can be up to 2.5, showing a large anisotropy. The observed anisotropy is well explained by the vortex motion in the context of type II superconductivity. Our findings have show-cased that superconductivity in InN films remains to be fully explored. Moreover, the zero-resistance state with a large critical current observed in the experiment indicates InN is a very interesting material for both fundamental research and the application on superconductor-semiconductor hybrid structure.

## Acknowledgements

The work in Rice was supported by NSF Grant DMR-0706634 and Welch Foundation Grant C-1682, in Peking University was supported by NSFC Grant No. 60990313 and No. 10774001, and RFDP Grant No. 20090001120008.

[*] E-mail address: rrd@rice.edu